\documentclass[aps,prl,twocolumn,showpacs,superscriptaddress,groupedaddress]{revtex4-1}  
\usepackage{graphicx}  
\usepackage{dcolumn}   
\usepackage{bm}        
\usepackage{amssymb}   
\usepackage{color}
\usepackage{placeins}
\usepackage{lineno,hyperref}

\begin{document}



\title{The role of the stochastic color field fluctuations on $J/\psi$ suppression in ultra-relativistic heavy-ion collisions} 
\date{\today}

\author{\it Ashik Ikbal Sheikh}
\address {Variable Energy Cyclotron Centre, HBNI, 1/AF Bidhan Nagar, Kolkata - 700064, India}
\author{\it Zubayer Ahammed}
\email{za@vecc.gov.in}
\address {Variable Energy Cyclotron Centre, HBNI, 1/AF Bidhan Nagar, Kolkata - 700064, India}
\author{\it Munshi  G.  Mustafa}
\address {Saha Institute of Nuclear Physics,  HBNI, 1/AF Bidhan Nagar, Kolkata - 700064, India}

\begin{abstract}
We consider the effect of the stochastic color field fluctuations in addition to the collisional as well
as the radiative energy losses in the propagation of charm quarks in the hot and dense deconfined
medium of quarks and gluons created in ultra-relativistic heavy-ion collisions. These fluctuations lead 
to energy gain of the propagating charm quarks.  We  construct and solve Langevin transport equations for charm quarks under the evolving background matter 
described by the ($3+1$)-D relativistic viscous hydrodynamics.  Considering the energy gain of charm quarks, the nuclear modification factor $R_{AA}$ of $J/\psi$ is calculated.
Interestingly, the experimental measurements for $J/\psi$ suppression in $Au-Au$ collisions at $\sqrt{s_{NN}} = 200$ GeV 
by PHENIX Collaboration and $Pb-Pb$ collisions at $\sqrt{s_{NN}} = 2.76$ TeV by ALICE 
and CMS Collaborations, can be nicely described with the effect of these field fluctuations without invoking the regeneration phenomena.

\end{abstract}

\maketitle


Quantum Chromodynamics (QCD) predicted formation of Quark-Gluon Plasma(QGP) in ultra-relativistic heavy-ion collisions~\cite{lqcd,qgp}. $J/\psi$ suppression due to the color screening effect has been considered as one of the most conclusive experimental evidence of the QGP formation. Several measurements of $J/\psi$ production have been reported over the last couple of years~\cite{sps,star,phenix,phenix11,aliceJHEP,alice,cms2tev,cms5tev}. However the results of $J/\psi$ production at LHC energies~\cite{aliceJHEP,alice,cms2tev,cms5tev} compelled intense discussions on the regeneration of  $J/\psi$~\cite{nu,pbm} apart from cold nuclear matter effects (CNM) \cite{cm}. While invoking these effects, other effects, importantly the role of chromo-electromagnetic field fluctuations were ignored.

The heavy quarks are mainly produced by hard scattering in the early stage of the ultra-relativistic heavy-ion collisions. After their production, they propagate through the dense medium and lose energy during their entire path of travel. This is reflected in the transverse momentum spectra and nuclear modification factor of heavy-flavour mesons. The magnitude of collisional energy loss is comparable to the radiative energy loss for heavy flavour quarks for certain domain of parton energies. Most of the studies estimated both radiative~\cite{mg99,mgm05,dokshit01,dead,wicks07,armesto2,B.Z,W.C,Vitev,AJMS,Saraswat15} and collisional~\cite{TG,BT,Alex,PP} energy loss of heavy quarks by considering the QGP medium without considering the microscopic field fluctuations. The parton energy loss due to stimulated gluon emission and thermal absorption is reported in Ref.~\cite{wang}. On the other hand, since QGP is a statistical system of dynamic colour charges, it can also be characterised by stochastic chromo-electromagnetic field fluctuations. These chromo-electromagnetic field fluctuations in the QGP causes an energy gain of heavy quarks of all momentum, significantly at the lower momentum limit~\cite{Fl}. 
This is due to the fact that the moving parton encounters the statistical change in energy in the QGP due to the fluctuations of the chromo-electromagnetic fields as well as the velocity of the particle 
under the influence of this field. It is essential to incorporate 
 these field fluctuations while describing the propagation of heavy quarks in QGP  as  shown in our earlier works~\cite{Ours1,Ours2}. The effects of such fluctuations were not considered earlier  while studying the $J/\psi$ production in heavy ion collisions.

In this Letter, for the first time,  the energy gain due to field fluctuations is  considered to the propagation of high energy charm quarks along with the energy loss caused by the collision and gluon radiation inside the QGP medium to study the $J/\psi$ propagation  in $Pb-Pb$ collisions at $\sqrt{s_{NN}} = 2.76$ TeV at the LHC. 
We find that the field fluctuations have substantial impact on $J/\psi$ suppression as observed in ALICE and CMS experiments~\cite{aliceJHEP,cms2tev}. The inclusion of the field fluctuations can describe the experimental observation of $J/\psi$ suppression without considering regeneration effects.

We consider charm quarks produced in the primordial hard scattering having initial momentum distribution  calculated up to  leading order (LO) with the centrality dependent nuclear parton distribution function EPS09~\cite{EPS09}. The  initial production points of charm quarks are distributed according to the nuclear overlap function of the colliding nuclei using Glauber model approach. For further details, we refer to the Ref.~\cite{Akamatsu}. For the evolution of the QGP medium, we follow (3+1)-D viscous hydrodynamics , vHLLE \cite{carpenko}. We use initial time  $\tau_{0} = 0.6$ $fm$, critical temperature $T_{c} = 150$ MeV. Shear viscosity $\eta/s = 0.08$ and bulk viscosity $\zeta/s = 0.04$ in the hadronic phase for $Au-Au$ and $Pb-Pb$ collisions are also used.  Optical Glauber initial state has been  used for space-time history of the flow velocity and temperature of the evolving medium. 
 
 With this initial phase space distribution of charm quarks, Langevin diffusion is performed under this hydrodynamic background,
 \begin{equation}
\label{xupdate}
dx_{i}  =  \frac{p_{i}}{E}dt,
\end{equation}

\begin{equation}
\label{pupdate}
dp_{i}  =  -\gamma p_{i}dt + \rho_{i} \sqrt {2Ddt},
\end{equation}
where $dx_{i}$ and $dp_{i}$ refer to the updates of the position and momentum of the charm quark in each time step $dt$ with $i = 1,\, 2$, and $3$ denotes the three components of position in Cartesian coordinates. We have used diagonal form for the diffusion matrix similar to earlier calculations~\cite{Hees,Cao,s17}. The standard Gaussian noise variable, $\rho_{i}$, is distributed randomly according to $w(\rho) =  \frac{1}{(2\pi)^{3/2}}\exp(-\rho^{2}/2)$ and  $\rho_{i}$ satisfies the relation, $<\rho_{i} > = 0$ and $ <\rho_{i}\rho_{j}> = \delta(t_{i}-t_{j}) $. The interactions between the charm quarks and the medium partons  are dictated by the drag ($\gamma$) and diffusion coefficient($D$ ). Following post-point discretization scheme, the equilibrium condition takes the form of fluctuation-dissipation theorem $D = \gamma E T$, where $E = \sqrt {p^2+M_{Q}^2}$ is the energy of the charm quark ($M_{Q}$ is the mass of the charm quark) and $T$
 is the temperature of the background medium. We have verified, in the large time limit, the charm quark phase space distribution function converges to the equilibrium 
 Boltzmann-J\'uttner  function  $e^{-E/T}$.
 
 In case of static medium  Eq.(\ref{xupdate}) and Eq.(\ref{pupdate}) are used. In our case( the evolving medium), we perform a Lorentz boost to each charm quark into the local rest frame of the fluid element through which it propagates and the position and momentum are updated according to Eq.(\ref{xupdate}) and Eq.(\ref{pupdate}).  Then  we boosted back to the laboratory rest frame by  performing inverse  Lorentz transformation to obtain the charm quark phase space coordinates. We stop the Langevin evolution when temperature of the background medium drops to  $150$ MeV, where particle spectra are calculated in statistical emission model~\cite{Chojnacki}. The hadronization of the charm quarks to $J/\psi$ is done by the Leading Order (LO) calculation of fragmentaion function as calculated in Ref.~\cite{Dadfar}.

Finally, we form the nuclear modification factor, defined as: 
  \begin{equation}
  R_{AA} =  \frac{dN^{AA}/dp_{T}}{N_{coll}dN^{pp}/dp_{T}}
  \end{equation}
  where, $N_{coll}$ is the number of binary nucleon-nucleon collisions for a given centrality class, obtained from Glauber model calculations.

\begin{figure} 
  \begin{minipage}{\columnwidth}
           \centering
            \includegraphics[width=\textwidth]{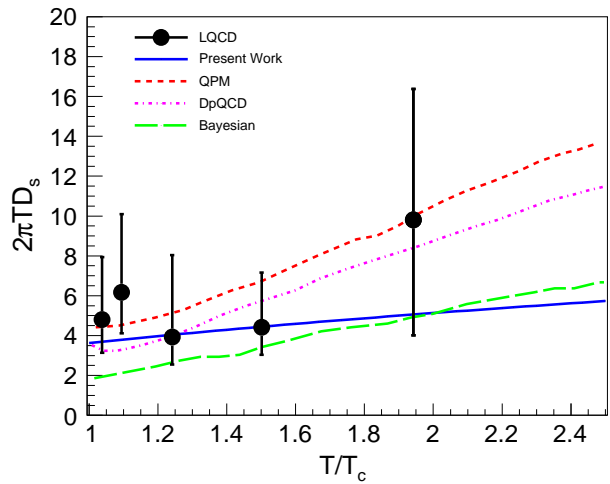}
            \caption{Spatial diffusion coefficient as a function of normalized temperature compared to lattice QCD calculations~\cite{lat}, dressed perturbative QCD(DpQCD), quasi-particle models(QPM)~\cite{qpart}, and Bayesian analysis~\cite{bay}. }
               \label{diff}
        \end{minipage}
\end{figure}

The  drag coefficient $\gamma$ for charm quark is calculated by using $\gamma = \frac{1}{p}(-\frac{dE}{dx})$\cite{MGM05,Santosh10}. Here we have considered both collisional\cite{PP} as well as radiative \cite{AJMS} energy loss. For the collisional energy loss , we have used modified differential energy loss calculations by Peigne and Pashier(PP)~\cite{PP} which is valid for the heavy quarks of all possible momenta. The radiative energy loss is calculated using the method discussed in Ref.\cite{AJMS,Saraswat15}. 
 This calculation of radiative energy loss considers the corrections for Dead Cone effect of heavy quarks and gives a very compact expression for gluon emission probability.
 Fig.\ref{diff} shows the dimensionless quantity 2$\pi$T$D_{s}$ ($D_{s}$ represents the charm quark spatial diffusion coefficient in the coordinate space) along with lattice calculations~\cite{lat}, Bayesian analysis~\cite{bay} and quasiparticle models(QPM)~\cite{qpart} . It is seen that our estimation of  2$\pi$T$D_{s}$ is consistent with other calculations.

Usually, the energy loss encountered by an energetic parton in a QGP medium reveals the dynamical properties of that medium in 
view of jet quenching of high energy partons. However, it is assumed that the collisional energy lost by the particle per unit time 
is small compared to the energy of the particle itself so that the change in the velocity of the particle during the motion may be neglected, 
i.e, the particle moves in a straight line trajectory~\cite{Fl}:. The energy loss of a particle is determined by the work of the retarding forces acting on 
the particle in the plasma from the chromo-electric field generated by the particle itself while moving. The collisional energy loss does not take into account the 
field fluctuations in the plasma and the particle recoil in collisions, which means that the medium is treated in an average manner, 
i.e., microscopic fluctuations were neglected~\cite{Fl}. Nevertheless,
the fluctuations of the chromo-electromagnetic field cause a statistical change in the 
energy of the moving charm quark inside plasma and the velocity of the charm quark under the influence of this field. As a consequence
of that, the charm quark gains energy and the leading log (LL) contribution of this gained energy is obtained by using semiclassical approximation as~\cite{Fl}:

\begin{eqnarray}
\left(\frac{dE}{dx}\right)_{\mbox{fl}}^{\mbox{LL}} = 2\pi C_F\alpha_{s}^{2}\left(1+\frac{n_f}{6}\right)\frac{T^3}{Ev^2}\ln{\frac{1+v}{1-v}} \ln{\frac{k_{\mbox{max}}}{k_{\mbox{min}}}},
\end{eqnarray}
where $k_{\mbox{max}} = \mbox{min}\left[E, {2q(E+p)}/{\sqrt{M^2+2q(E+p)}} \right]$ with
$q \sim T$ is the typical momentum of the thermal partons (light quarks and gluons) in the QGP and $k_{\mbox{min}} = \mu_g $ is the Debye mass. 
 It is to be noted that the semiclassical approximation is equivalent to the hard thermal loop approximation based on weak coupling limit. This gained energy is taken into account while performing the position and momentum updates in the Langevin diffusion process of charm quarks.
 
\begin{figure} 
 	\begin{minipage}{\columnwidth}
 		\centering
 		\includegraphics[width=\textwidth]{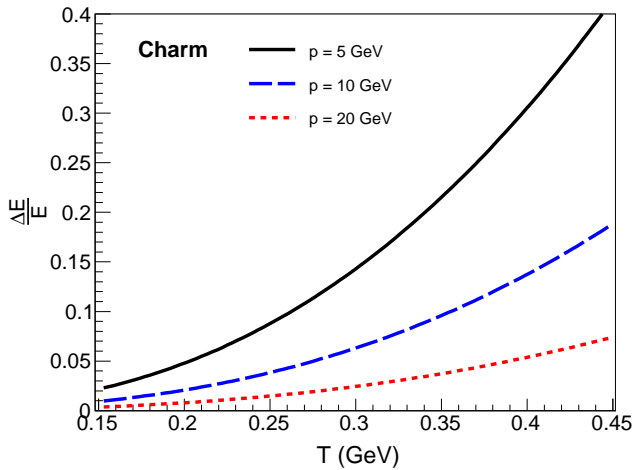}
 		\caption{Fractional energy gain of a charm quark with momentum, $p=5$, $p=10$ and $p=20$ GeV inside the QGP medium due to fluctuations, as a function of $T$. The path length considered here is $L = 5$ fm.
 			 }
 		\label{fracEgain}
 	\end{minipage}
 \end{figure}

 Fig.\ref{fracEgain} displays the fractional energy gain of a charm quark with momentum, $p=5$, $p=10$ and $p=20$ GeV inside the QGP medium due to field fluctuations as a function of $T$. The fractional energy gain is more pronounced at lower momenta. 
 
\begin{figure}
  \begin{minipage}{\columnwidth}
           \centering
            \includegraphics[width=\textwidth]{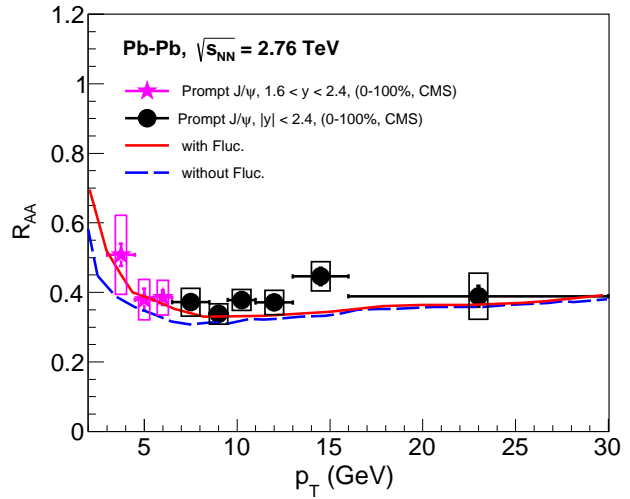}
            \caption{The nuclear modification factor $R_{AA}$ of $J/\psi$ mesons with the effect of fluctuations as a function of $p_{T}$ in $Pb-Pb$ collisions for $0-100$$\%$ centrality at $\sqrt{s_{NN}} = 2.76$ TeV. The experimental measurements are taken from CMS Collaboration~\cite{cms2tev}.}
               \label{raaminbias}
        \end{minipage}
\end{figure}
\begin{figure} 
  \begin{minipage}{\columnwidth}
           \centering
            \includegraphics[width=\textwidth]{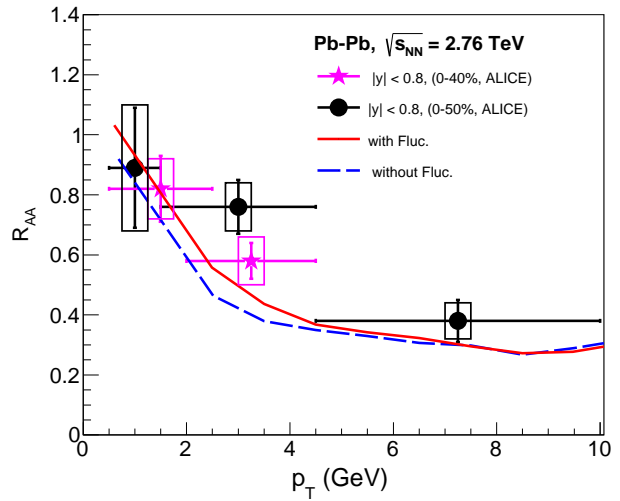}
            \caption{The nuclear modification factor $R_{AA}$ of $J/\psi$ mesons with the effect of fluctuations as a function of $p_{T}$ in $Pb-Pb$ collisions at $\sqrt{s_{NN}} = 2.76$ TeV, compared with the ALICE data~\cite{aliceJHEP}.}
               \label{raac050}
        \end{minipage}
\end{figure}

\begin{figure} 
  \begin{minipage}{\columnwidth}
           \centering
            \includegraphics[width=\textwidth]{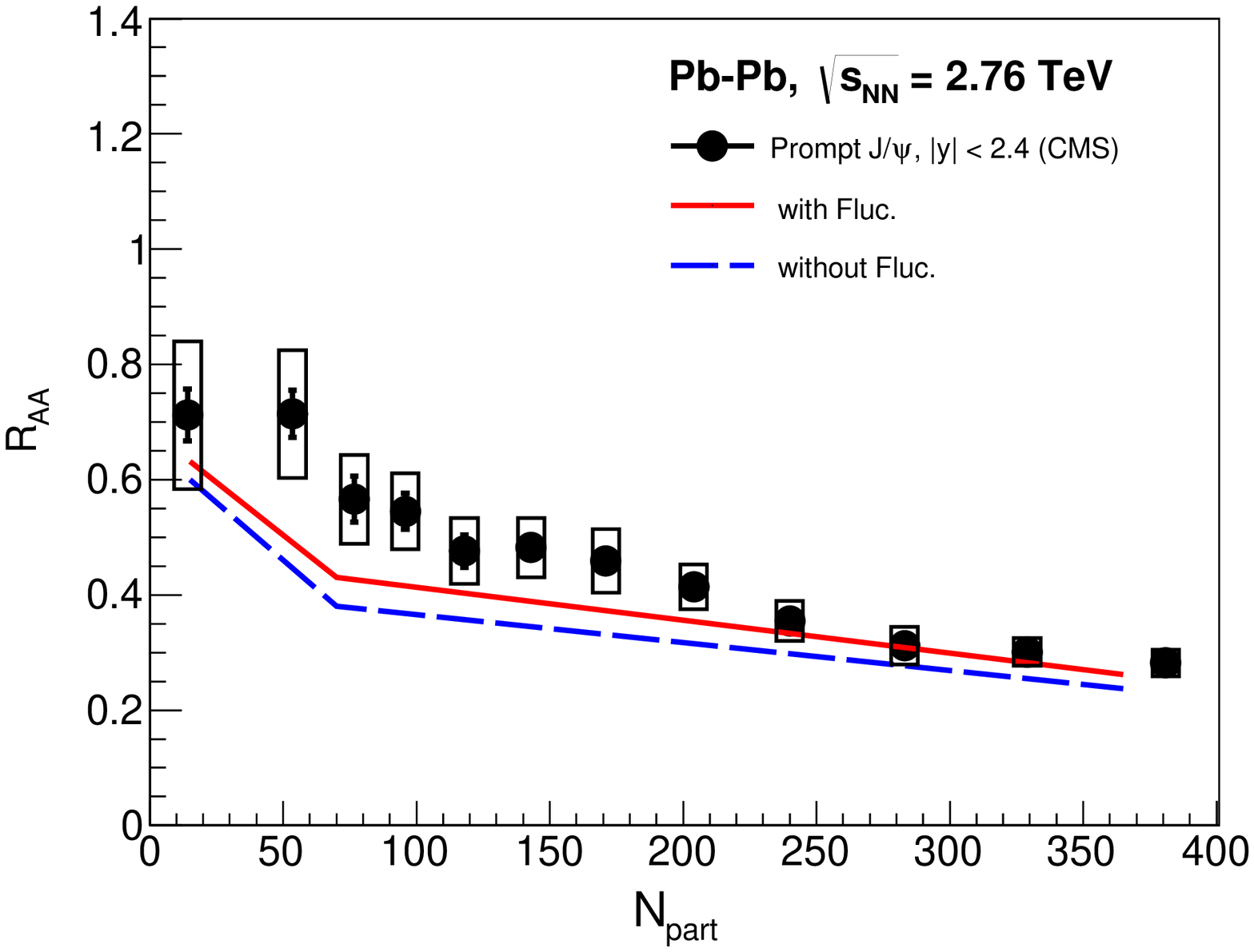}
            \caption{The nuclear modification factor $R_{AA}$ of $J/\psi$ mesons with the effect of fluctuations as a function of $N_{part}$ in $Pb-Pb$ collisions at $\sqrt{s_{NN}} = 2.76$ TeV. The experimental measurements are taken from CMS Collaboration~\cite{cms2tev}.}
               \label{raanpartlhc}
        \end{minipage}
\end{figure}
\begin{figure} 
  \begin{minipage}{\columnwidth}
           \centering
            \includegraphics[width=\textwidth]{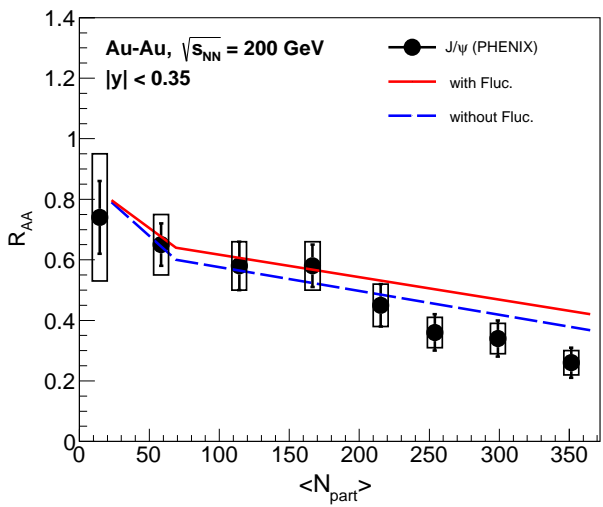}
            \caption{The nuclear modification factor $R_{AA}$ of $J/\psi$ mesons with the effect of fluctuations as a function of $N_{part}$ in $Pb-Pb$ collisions at $\sqrt{s_{NN}} = 2.76$ TeV. The experimental data are taken from PHENIX Collaboration~\cite{phenix}.}
               \label{raanpartrhic}
        \end{minipage}
\end{figure}

The nuclear modification factor, $R_{AA}$ for $J/\psi$ as a function of transverse momentum $p_{T}$ is shown in Fig.\ref{raaminbias} with and without considering chromo-electromagnetic field fluctuations. Significant contribution of chromo-electromagnetic field fluctuations is observed at lower transverse momentum region. The estimated $R_{AA}$ is in good agreement with the measured experimental results \cite{cms2tev} by CMS collaboration. In Fig.\ref{raac050}, we show  the results for $0-40\%$ and $0-50\%$ centrality class data as measured by ALICE collaboration at mid-rapidity~\cite{aliceJHEP}. The estimated $R_{AA}$ with chromo-electromagnetic field fluctuations reproduces the  measured experimental data. The energy gain due to field fluctuations occurs as gluons are absorbed during their propagation. This energy gain due to field fluctuations is more prominent in the low $p_T$ region. This is consistent with the observed comparison with the data.
We show the numerical values of $R_{AA}$  as a function of number of participant nucleons in Fig.\ref{raanpartlhc}.The results from CMS experiment is also compared. It is observed that the estimated values of $R_{AA}$ with field  fluctuations are in good agreement with the data within their uncertainties.
We  have also carried out similar studies at RHIC top energy. Fig.\ref{raanpartrhic} displays the calculated values of $R_{AA}$ for Au+Au collisions at $\sqrt{s_{NN}} = 200$ GeV. The results are compared with the data of PHENIX collaboration~\cite{phenix}. We observe that the experimental data can be described  without considering chromo-electromagnetic field fluctuations which indicates no regeneration of  $J/\psi$  is necessary in the RHIC energy.
 Nevertheless, we also show the $J/\psi$ suppression with field fluctuations. It is to be noted here that the semiclassical approximation equivalent to the hard thermal loop
  approximation in the weak coupling limit is used to calculate mean energy loss and energy gain due to field fluctuations. 
  The uncertainties in the used fragmentation function  may also include uncertainties in our calculations.

In summary, we have performed the Langevin diffusion of charm quarks in a evolving hydrodynamic background medium with the effect of the chromo-electromagnetic field fluctuations and studied $J/\psi$ suppression in heavy-ion collisions  at $\sqrt{s_{NN}} = 2.76$ TeV. The effect of the field fluctuations leads to energy gain of charm quarks. The chromo-electromagnetic field fluctuations play an important role and probably one can closely explain the measured experimental data at $\sqrt{s_{NN}} = 200$ GeV at RHIC and $\sqrt{s_{NN}} = 2.76$ TeV  at LHC energy without invoking regeneration phenomena.

We are very much grateful to Yuriy Karpenko for providing hydrodynamic outputs. We would like to thank  D.K. Srivastava, Santosh K. Das and Jane Alam for fruitful discussions and suggestions.

\end{document}